\documentclass{PoS}
\usepackage{nicefrac}
\usepackage{amsmath}

\title{The New Muon $g-2$ Experiment at Fermilab}

\ShortTitle{The New Muon $g-2$ Experiment at Fermilab}

\author{\speaker{Joseph Grange}\\
        Argonne National Laboratory\\
        E-mail: \email{grange@anl.gov}}

\abstract{Precision measurements of fundamental quantities have played a key role in pointing the way forward in developing our understanding of the universe.  Though the enormously successful Standard Model (SM) describes the breadth of both historical and modern experimental particle physics data, it is necessarily incomplete.  The muon $g-2$ experiment executed at Brookhaven concluded in 2001 and measured a discrepancy of more than three standard deviations compared to the Standard Model (SM) calculation~\cite{821}.  Arguably, this remains the strongest hint of physics beyond the SM.  A new initiative at Fermilab is under construction to improve the experimental accuracy four-fold.  The current status is presented here.}

\FullConference{16th International Workshop on Neutrino Factories and Future Neutrino Beam Facilities - NUFACT2014,\\
		25 -30 August, 2014\\
		University of Glasgow, United Kingdom }

\begin{document}

\section{Introduction and $g-2$ theoretical value}

Any spin-1/2 charged particle has an intrinsic magnetic moment $\vec{\mu}$ that is aligned with its spin vector $\vec{s}$:

\begin{center}
\begin{eqnarray}
\label{eqn:magmom}
\vec{\mu} = g\left(\frac{q}{2m}\right)\vec{s},
\end{eqnarray}
\par\end{center}

\noindent where $q (m)$ is the particle's charge (mass), and $g$ is the gyromagnetic factor discussed here.

\begin{figure}
\begin{center}$
\begin{array}{ccc}
  \includegraphics[scale=0.15]{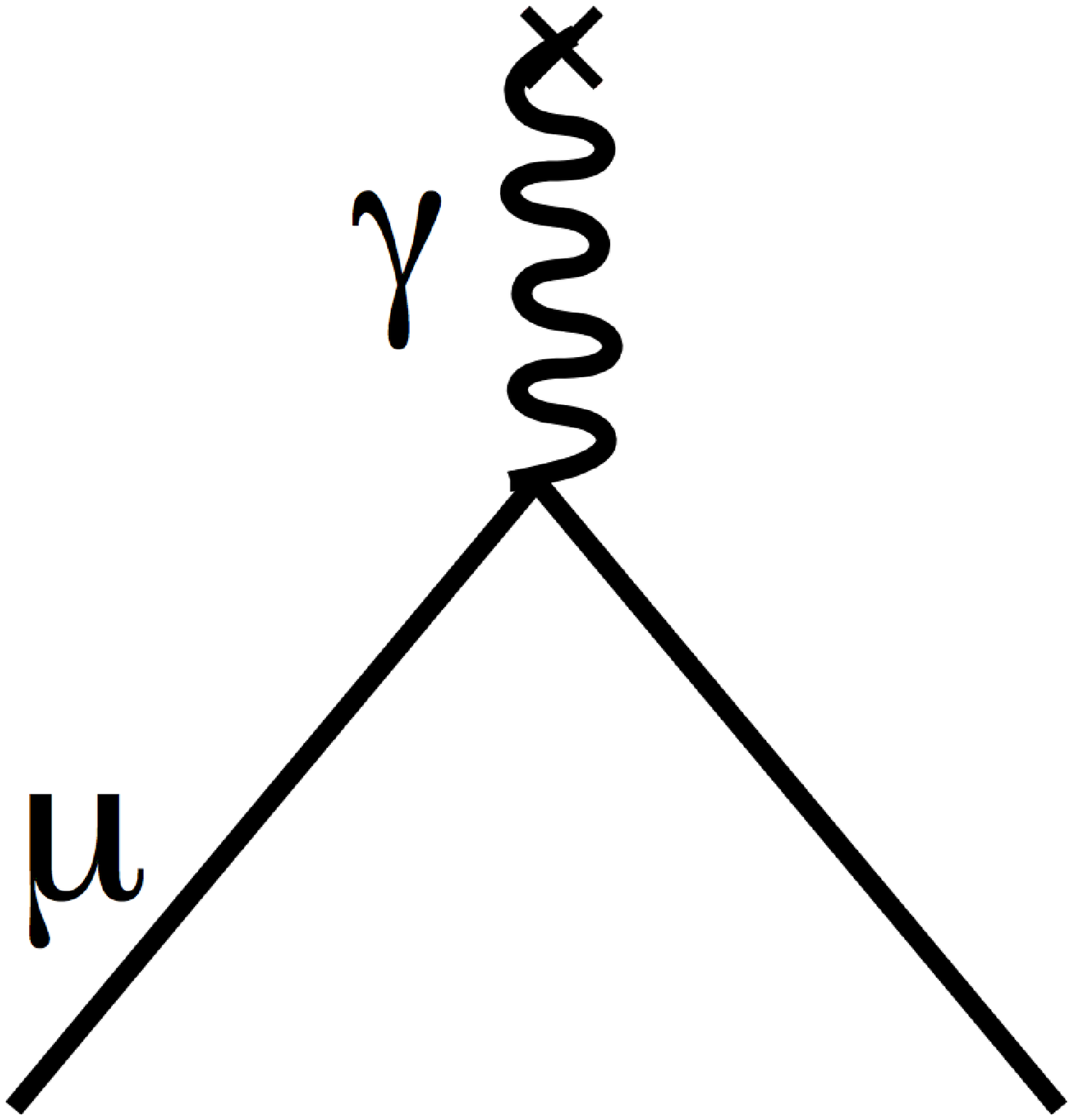} & \hspace{2cm}
\includegraphics[scale=0.15]{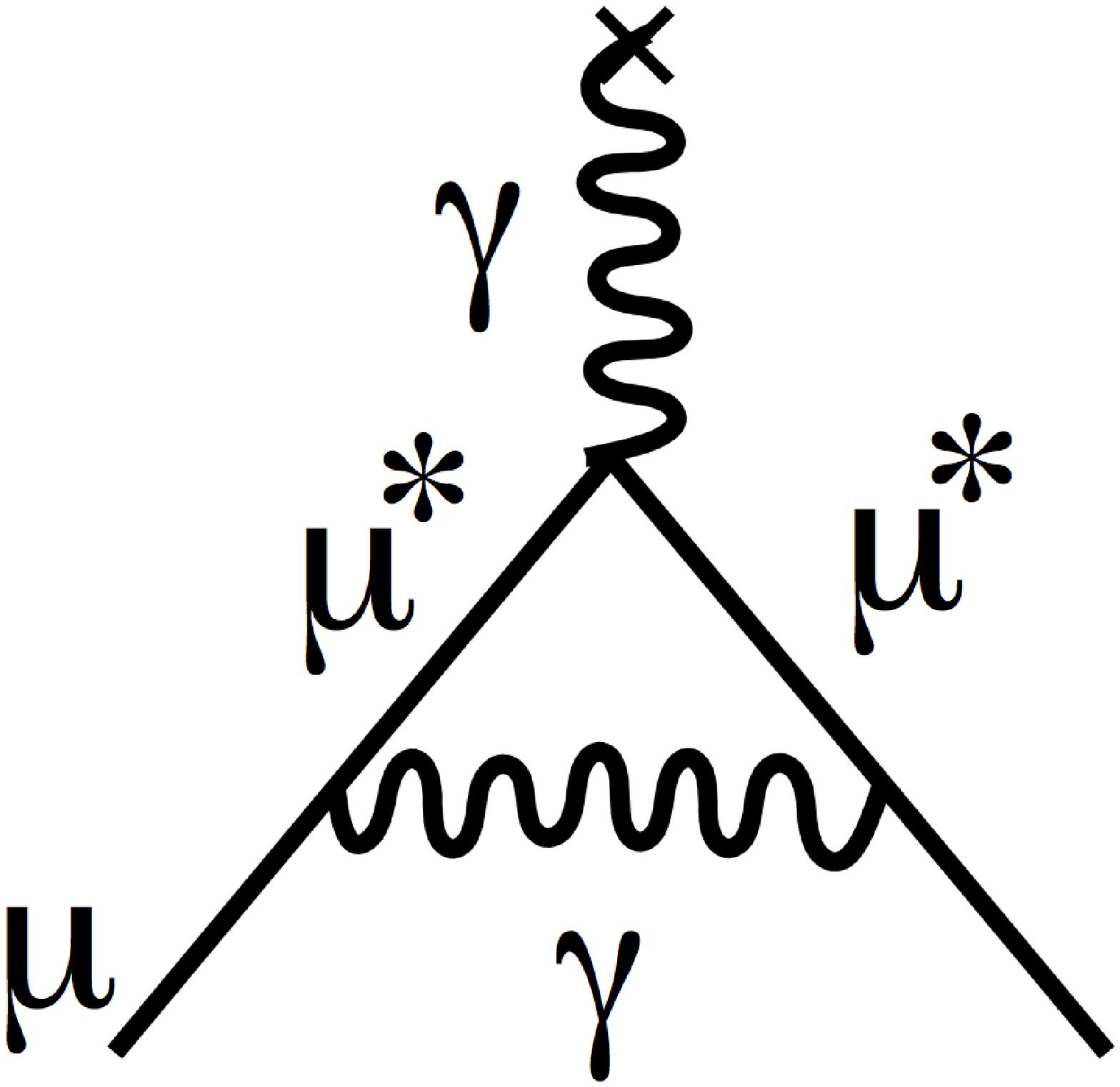} & \hspace{1.8cm}
\includegraphics[scale=0.20]{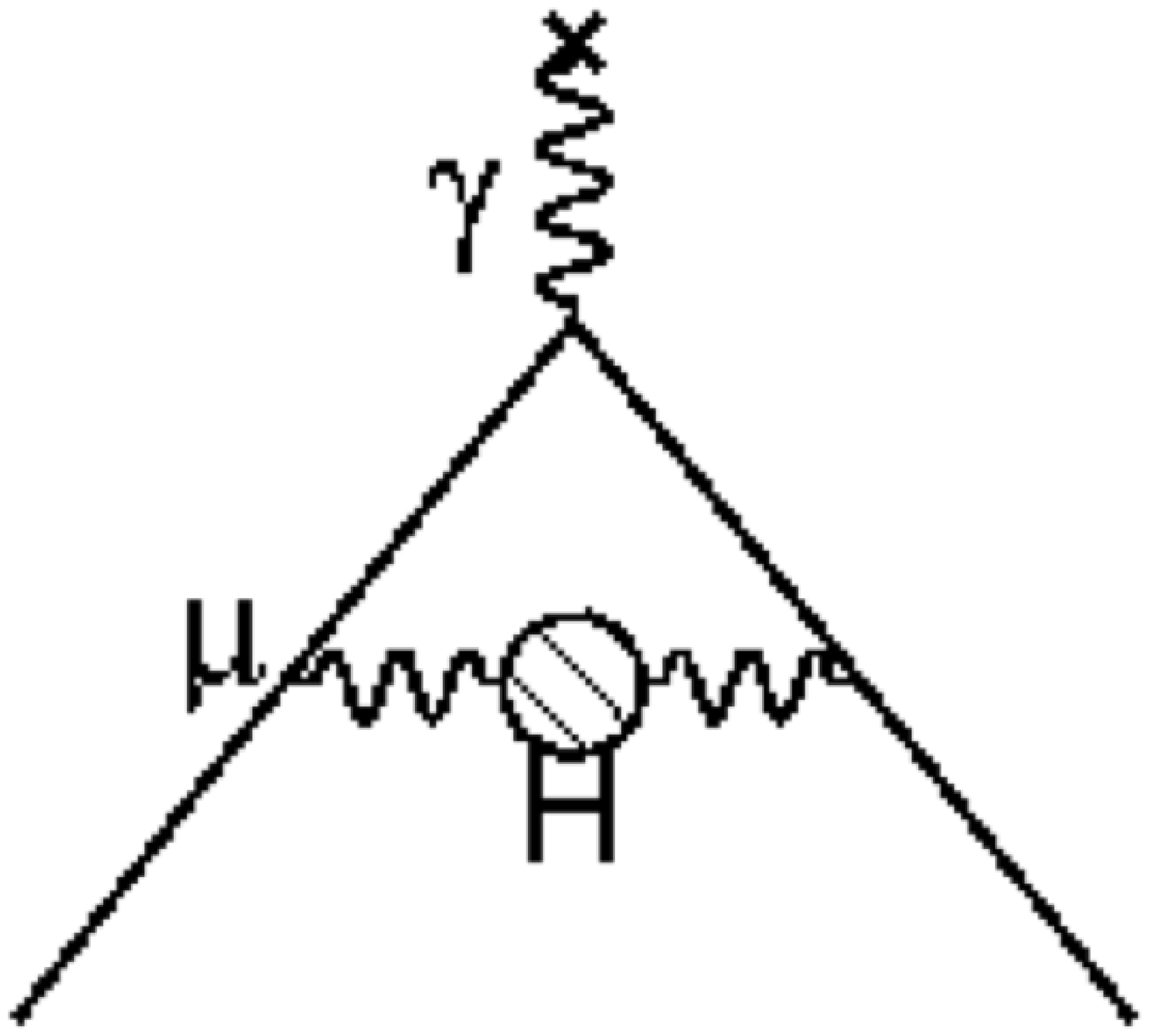} \\
  {\bf (a)}&\hspace{2cm} {\bf (b)}& \hspace{1.8cm}{\bf (c)}\\
\end{array}$
\end{center}
\caption{Feynmann diagrams for g=2 (a), the Schwinger term (b), and the first-order hadronic vacuum polarization contribution (c).}
\label{fig:feyn} 
\end{figure}

During the birth of quantum mechanics, P. A. M. Dirac predicted $g=2$ for non-composite particles~\cite{dirac}.  As an example of the insight afforded by his description, measurements of $\sim 5.6$ and $-3.8$ for the proton and neutron magnetic moments, respectively, yielded the first evidence of nucleon substructure.  However, while accurate in general to better than 1\% for fundamental particles, the theory was incomplete.  A large discrepancy between its prediction and a measurement of the hydrogen hyperfine structure some decades later~\cite{hhs} prompted work by J. Schwinger that ultimately resolved this disagreement and initiated the field of QED~\cite{schwinger}.  The interactions between virtual particles and the real particle perturb its gyromagnetic factor $g$ at the $\sim$ 0.1\% level for electrons and muons.  The perturbation is dominantly due to electromagnetic single photon exchange (Figure~\ref{fig:feyn}(b)), but the exact value is sensitive to all such allowed interactions.  The measurement and subsequent explanation of the magnetic moment of various particles has already significantly reduced our ignorance of the quantum world, and so tests of this quantity with increasing precision are well-motivated to search for physics beyond the SM.  The muon $g-2$ value was measured most recently at Brookhaven National Laboratory (BNL), where a discrepancy of more than three standard deviations was found~\cite{821}.  A new experiment at Fermilab aims to perform a measurement four times more accurate to determine if this disagreement is due to new physics.

The first order QED term, so-called the Schwinger term, $\nicefrac{\alpha}{2\pi}\sim 10^{-3}$ (Figure~\ref{fig:feyn}(b)) dominates the deviation of the magnetic moment from two.  The overall QED contribution to $g$ has been calculated with high precision up to 10$^{\textrm{th}}$ order interactions~\cite{kino}.  It is the hadronic diagrams (Figure~\ref{fig:feyn}(c)) that dominate the overall theory uncertainty~\cite{hads}, as parts of its evaluation are based on experimental data or is model dependent.  The current status of the muon $g-2$ theoretical value contributions is summarized in Table~\ref{tbl:SMvalue}, where the anomalous moment is recast into units of $a_\mu\equiv\nicefrac{g-2}{2}$.  The total uncertainty on this prediction is comparable in size with the recent muon $g-2$ experimental uncertainty.  Therefore, future experimental improvements must be accompanied by simultaneous theoretical uncertainty reductions to maximize the sensitivity to new physics.  On the timescale of the experiment, the theory is expected to improve by $\sim$30\% compared to the current uncertainty~\cite{tdr}.

\begin{table}[h!]
\begin{center}
\caption{Summary of the Standard-Model contributions to the muon anomaly $a_\mu\equiv\nicefrac{g-2}{2}$.  The lowest order (lo) of hadronic vacuum polarization (HVP) contribution is limited by experimental data, while the hadronic light-by-light (HLbL) term is heavily model dependent.  Both of these terms are expected to be known more accurately on the timescale of the Fermilab muon $g-2$ experimental results.}
{\small
\begin{tabular}{lr}
\hline \hline {\sc\small  }Contribution &
{\sc\small  $a_\mu (\times \, 10^{-11})$ units}
\\ \hline
QED ($\gamma + \ell$)~\cite{kino} & $116\,584\,718.951\pm 0.077$ \\
HVP(lo)~\cite{hads} & $6\,949 \pm 43$\\
HLbL~\cite{glasgowcon} &  $105\pm 26 $ \\
EW~\cite{EW} & $153.6\pm 1.0 $ \\
 \hline
 Total SM~\cite{hads} & $116\,591\,802 \pm 42_{\mbox{\rm \tiny H-LO}} 
\pm 26_{\mbox{\rm \tiny H-HO}}  \pm 2_{\mbox{\rm \tiny other}}   \, (\pm 49_{\mbox{\rm \tiny tot}}) $\\
 \hline\hline\
\end{tabular}}
\label{tbl:SMvalue}
\end{center}
\end{table}

\section{Experimental overview}

Under the influence of a magnetic field $\vec{B}$, a muon will exhibit two important behaviors.  Kinematically, its momentum will rotate ($\nicefrac{d\vec{p}}{dt} = e\vec{v}\times\vec{B}$):

\begin{center}
\begin{eqnarray}
\label{eqn:cyclo}
\omega_c = \frac{eB}{\gamma m c},
\end{eqnarray}
\par\end{center}
 
\noindent where $\omega_c$ is the cyclotron frequency.  Continual momentum rotation allows for muon storage.  The second important behavior is the Larmor precession of its spin vector due to torque provided by the magnetic field ($\nicefrac{d\vec{s}}{dt}=\vec{\mu}\times\vec{B}$):

\begin{center}
\begin{eqnarray}
\label{eqn:spin}
\omega_s = \frac{g_\mu e B}{2mc} + (1-\gamma)\frac{e B}{\gamma m c},
\end{eqnarray}
\par\end{center}

\noindent where $\omega_s$ is the spin rotation frequency.  As is the case for the $g-2$ experiment, the second term in $\omega_s$ (Thomas precession) is appropriate for a rotating inertial system.  Assuming the velocity is entirely perpendicular to the magnetic field, the measurable difference of the two quantities $\omega_s$ and $\omega_c$ is directly sensitive to the muon anomaly $a_\mu$:

\begin{center}
\begin{eqnarray}
\label{eqn:amu}
\omega_a \equiv \omega_s - \omega_c = a_\mu\frac{qB}{m}.
\end{eqnarray}
\par\end{center}

In the present experimental context where the source of naturally polarized muons originate from decay-in-flight pions, a static electric field inside the storage ring is employed to enforce vertical containment.  This modifies the precession of the anomalous magnetic moment frequency (Eq.~\ref{eqn:amu}) to read:

\begin{center}
\begin{eqnarray}
\label{eqn:amufull}
\omega_a =-\frac{q}{m}\left( a_\mu B - \left[a_\mu - \frac{1}{\gamma^2-1} \right]\frac{\lvert\vec{\beta}\times \vec{E}\rvert}{c} \right).
\end{eqnarray}
\par\end{center}

\begin{figure}
\begin{center}$
\begin{array}{c}
\includegraphics[scale=0.50]{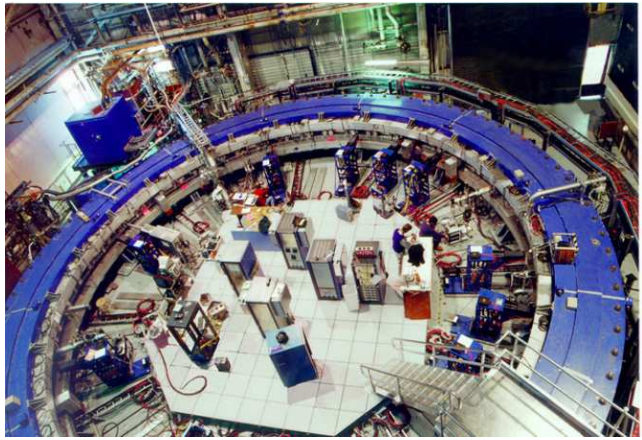} \\
\end{array}$
\end{center}
\caption{The $g-2$ muon storage ring at BNL.}
\label{fig:ring} 
\end{figure}

If comparable in size to the first parenthetical term $a_\mu B$, the second term above would imply large experimental sensitivity to beam dynamics and $\vec{E}$ field in the determination of $a_\mu$.  Fortunately, a judicious choice of $\gamma$ minimizes its contribution.  Using $\gamma = \sqrt{\nicefrac{\left(1+a_\mu\right)}{a_\mu}} \sim 29.3$ ($p_\mu=3.1$ GeV/c) reduces the effective contribution to a small and well-determined correction~\cite{tdr}.  This choice of energy determines the relationship between the magnetic field strength and the storage ring radius where the anomalous magnetic moment frequency is observed.  The BNL muon $g-2$ experiment produced a 7.1~m radius storage ring permeated by a highly homogeneous magnetic field of strength 1.45~T.  Figure~\ref{fig:ring} shows a photo of the ring at BNL.  It is currently being reassembled at Fermilab.

By exploiting the experience and wealth of experimental techniques used in the E821 measurement, the task ahead of the Fermilab $g-2$ experiment is to measure the muon spin precession frequency and the magnetic field even more accurately while collecting an unprecedented amount of statistics.  Table~\ref{tbl:uncoview} shows an overview of the statistical and systematic uncertainties for the BNL measurement and projections for the Fermilab determination.

\begin{table}[h!]
\begin{center}
\caption{Summary of experimental uncertainties for the BNL muon $g-2$ measurement and projections for the new experiment at Fermilab.  Fractional uncertainties on the quantity $a_\mu$ = $\nicefrac{g-2}{2}$ are given in units of parts per billion (ppb).}
{\small
\begin{tabular}{ccc}
\hline 
\hline 
$a_\mu$ uncertainty source & BNL (ppb)~\cite{821} & FNAL goal (ppb)~\cite{tdr}\\ 
\hline
$\omega_a$ statistics & 480 & 100 \\
$\omega_a$ systematics & 180 & 70 \\
$\omega_p$ systematics & 170 & 70 \\
\hline
 {\bf Total} & {\bf 540} & {\bf 140} \\
 \hline\hline\
\end{tabular}}
\label{tbl:uncoview}
\end{center}
\end{table}

\section{Magnetic field production and measurement}

\begin{figure}
\begin{center}$
\begin{array}{cc}
\includegraphics[scale=0.22]{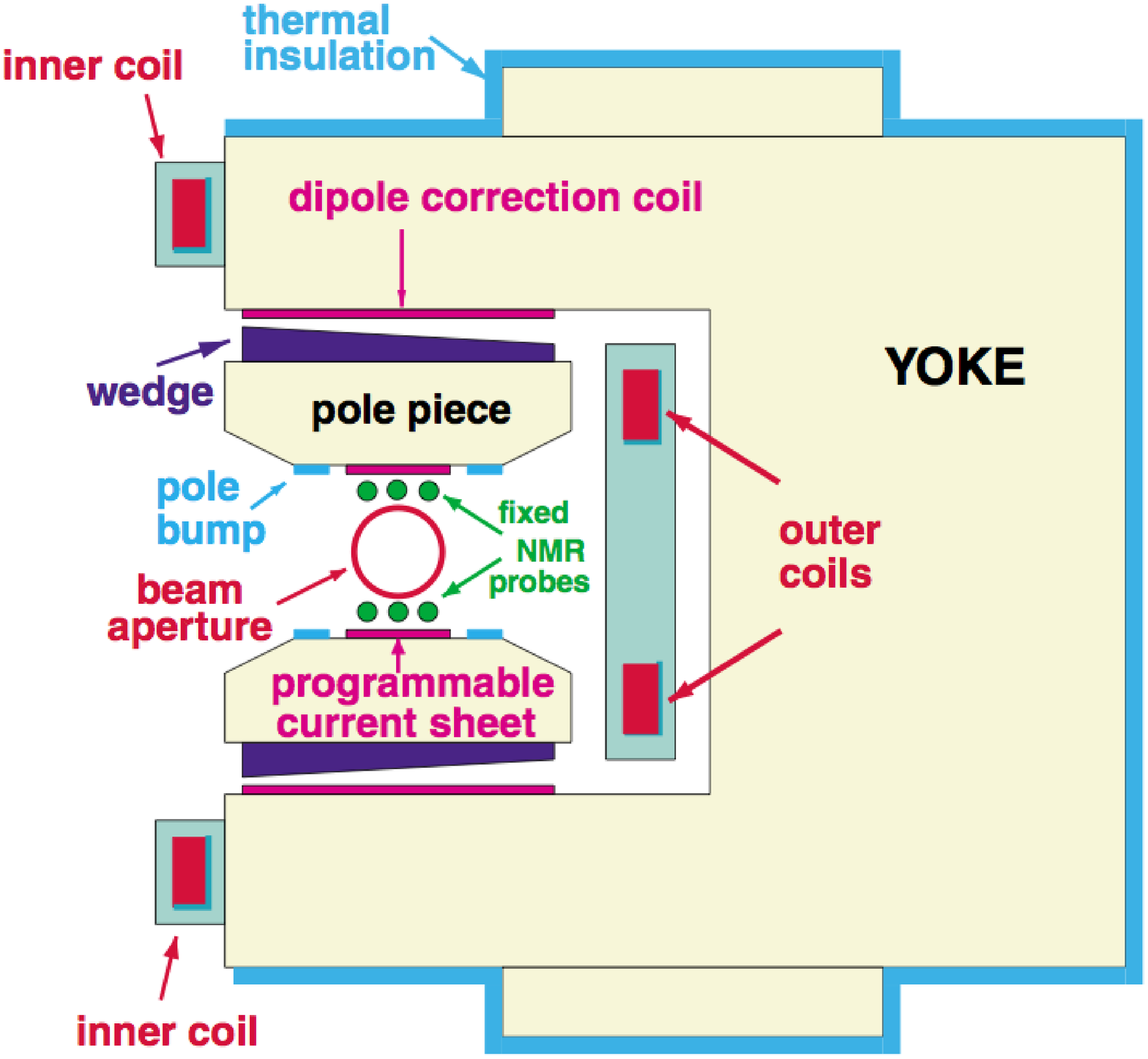} &
\includegraphics[scale=0.25,trim=-150 -85 0 0]{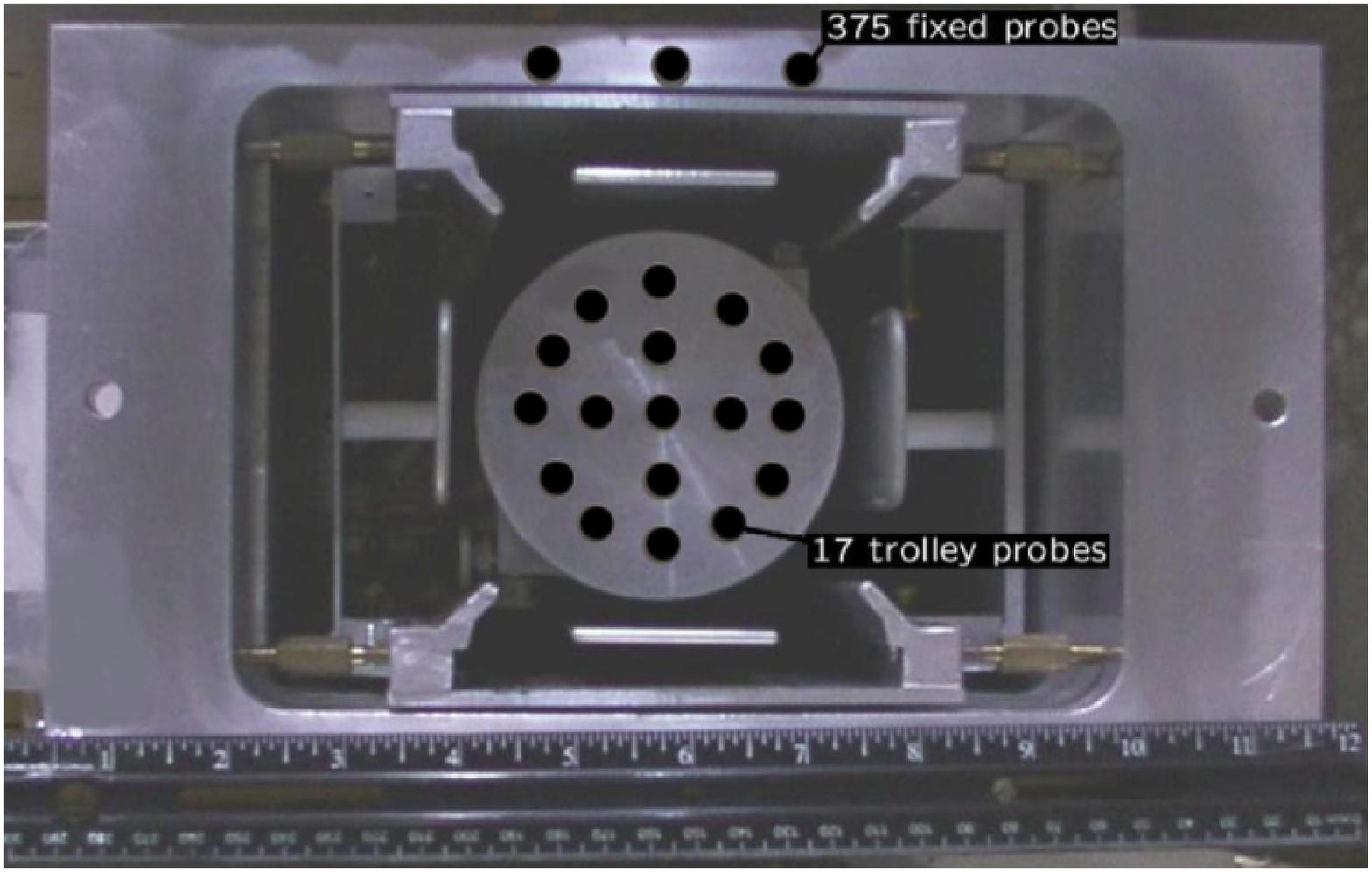} \\
\end{array}$
\end{center}
\caption{Cross section of the storage ring (left) with the various shimming tools and cross section of a vacuum chamber (right) showing the NMR magnetometers.}
\label{fig:shim} 
\end{figure}

Superconducting coils in the $g-2$ experiment carry $\sim$ 5200~A to excite a 1.45~T magnetic field in the storage region. The magnetic field referenced in Eq.~\ref{eqn:amu} in practice is the average field experienced by the stored muon distribution.  Inevitably, beam dynamics and muon momentum spread lead to a spatial sampling of the storage ring.  Therefore, a magnetic field homogeneous in both the longitudinal and transverse dimensions minimizes the need for detailed knowledge of beam dynamics and local field gradients.  As shown in Figure~\ref{fig:shim}, a large number of shimming tools that control various aspects of the field is integrated into the surrounding steel for this purpose.  

The tools to measure this field are shown in Figure~\ref{fig:shim}, where a cross-section of the vacuum chamber shows the positions of the tools used to measure the resultant field. These tools are water- or petroleum jelly-based nuclear magnetic resonance (NMR) devices~\cite{nmr}.  As indicated, almost 400 NMR probes are stationary around the circumference in the ring both on top and on bottom of the vacuum chambers, while a mobile ``trolley" device is regularly deployed to directly measure the field distribution in the storage volume during beam-off conditions.

While the NMR probes used to measure the absolute magnetic field are typically precise to $\sim20$ parts per billion (ppb), diamagnetic shielding inside the sample volume shift the accuracy by 10's of parts per million (ppm).  Therefore, a precise calibration must be performed with a special NMR probe and this must then be carefully transferred to the probes used in the experiment.  Various aspects of this delicate procedure dominated the systematic uncertainty of the field measurement for the BNL experiment~\cite{821}.

In E821 the measurement was limited by statistics.  So that the new experiment is not limited by systematics, many of the systematic uncertainties require more careful evaluation~\cite{tdr}.


\section{Spin precession measurement}

\begin{figure}
\begin{center}$
\begin{array}{cc}
\includegraphics[scale=0.3]{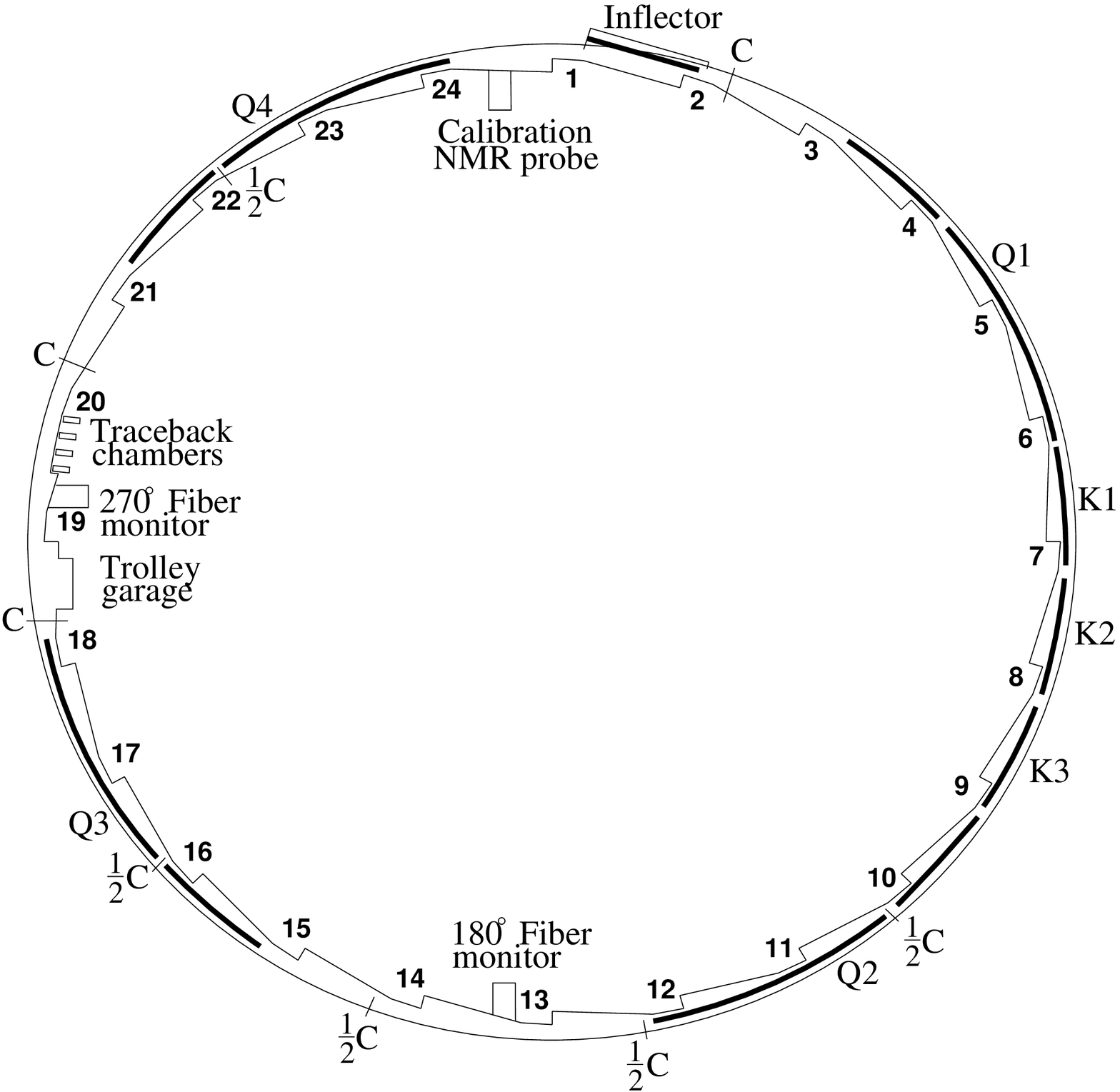}&
\includegraphics[scale=0.45,trim=0 -100 0 0]{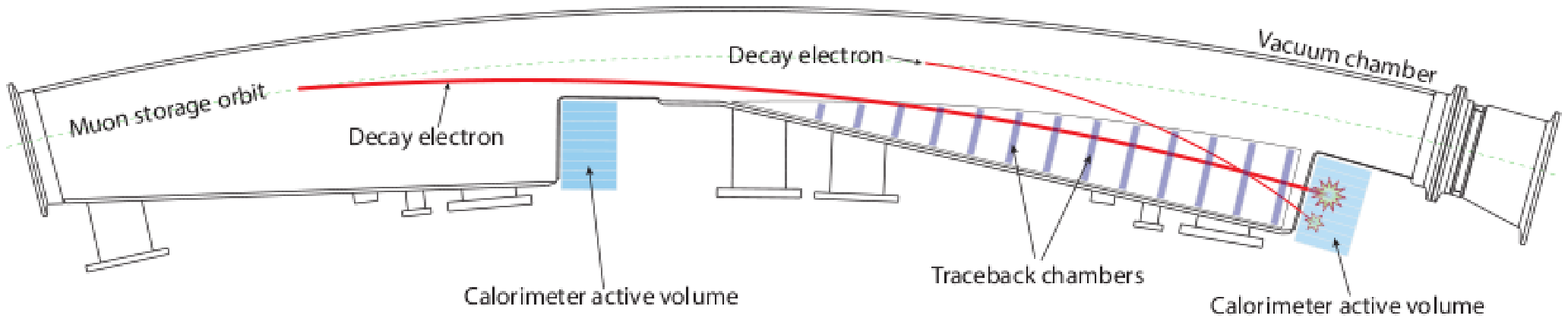}\\
\end{array}$
\end{center}
\caption{Overhead view of the $g-2$ storage ring as used in the BNL measurement (left) and details of electron detection inside a vacuum chamber (right) including one of the new in-vacuum tracker detectors.  The vacuum chambers feature a scalloped shape to minimize multiple scattering prior to detection.  Calorimeter stations are marked in the left figure by numbers 1-24.  The in-air tracker chamber just upstream of calorimeter station 20 will be replaced by three in-vacuum systems upstream of stations 15, 21 and 3.  Also shown are the beam storage and monitoring systems: the inflector~\cite{inf}, electric quadrupole stations (labeled by ``Q")~\cite{quadnim}, collimators (``C"), fiber beam monitors, and kicker magnets (``K")~\cite{kicker}.}
\label{fig:overheadring} 
\end{figure}

Recall the self-analyzing weak decay of the muon ($\mu \rightarrow e \nu \bar{\nu}$): the combination of parity violation with conservation of spin and lepton number results in a correlation between the emitted electron direction and the muon spin direction at the time of decay.  Moreover, this correlation is dependent on the electron energy and is maximal for the highest energy decay electrons.  The anomalous precession frequency $\omega_a$ (Eq.~\ref{eqn:amu}) is then measured by observing the kinematic precession of high-energy ($> 1.8$ GeV) electrons with a set of 24 segmented and high-density (PbF$_2$) calorimeters symmetrically distributed on the ring's interior.  Figure~\ref{fig:overheadring} shows an overview of the important hardware components inside and outside of the storage volume as implemented in the Brookhaven experiment and also a cartoon of electron detection inside a vacuum chamber.  

To recover the amplitude and phase of the $g-2$ precession based on observations of electron kinematics, the new calorimeters must provide good energy and excellent timing resolution.  These calorimeters have been thoroughly tested and characterized with an electron beam at SLAC~\cite{calos}, and results indicate their performance will meet the physics requirements.  The final configuration will also feature spatial segmentation to reduce pileup and deduce the mean vertical position of the muon beam.

Figure~\ref{fig:overheadring} also describes the placement of in-vacuum tracking chambers, which will be located at three locations around the storage ring.  These nearly massless detectors ($\sim$10$^{-3}$ X$_0$)  sit just upstream of the calorimeters and will provide further pileup resolution as well as beam dynamics information that will allow data-based constraints on many beam-related systematic uncertainties~\cite{tdr}.  

The precession vector $\vec{\omega}_a$ will be observed for roughly ten muon lifetimes for each fill ($\sim 700 \mu$s).  Any systematic change in the detection of electron kinematics across this timeframe directly biases the $\vec{\omega}_a$ measurement.  One such effect is provided by the stability of the calorimeter gain, where it will observe rates differing by four orders of magnitude over less than 1~ms.  A laser system will be deployed to inject photons into each calorimeter channel and will be compared to the output signals to provide direct energy calibration measurements. To reduce the uncertainty induced by calorimeter gain changes to a negligible level, a sub-per mil accuracy on the measured electron kinematics must be achieved.  This requirement is at least an order of magnitude more precise than any previous particle physics based calibration system measurement for the purposes of calorimetry.

\section{Summary}

Many important breakthroughs in the history of particle physics have been incited by precision measurements of fundamental quantities.  The BNL muon $g-2$ experiment observed a discrepancy of more than three standard deviations when compared to the SM prediction.  This measurement was limited by the collected muon precession statistics and not the technology, so an improved measurement to determine whether the BNL data is a result of new physics or an unlikely statistical fluctuation is compelling.

Though the BNL $g-2$ measurement was not limited by hardware or analysis techniques, the systematic uncertainties encountered there would preclude a dramatic improvement in accuracy if the sole change were increased statistics.  Therefore, greater than a two-fold reduction in systematic uncertainties is also planned and poses an important challenge in the coming years.

The Fermilab muon $g-2$ experiment is scheduled to begin collecting production data in early 2017.

\end{document}